\documentclass{iau}
\usepackage{graphicx}
\usepackage{natbib}
\usepackage{aas_macros}
\usepackage{multicol}

\title[The Physics of Galaxy Evolution with SPICA] 
{The physics of Galaxy Evolution with SPICA observations}

\author[Spinoglio, Fern\'andez-Ontiveros \& Mordini]   
{Luigi Spinoglio, 
 Juan A. Fern\'andez-Ontiveros \and Sabrina Mordini}

\affiliation{
Istituto di Astrofisica e Planetologia Spaziali - INAF, Rome, \\ Via Fosso del Cavaliere 100, 00133, Roma, Italia\\
email: {\tt luigi.spinoglio@iaps.inaf.it, j.a.fernandez.ontiveros@gmail.com, sabrina.mordini@uniroma1.it} \\[\affilskip]
}

\pubyear{2020}
\volume{359}  
\jname{Galaxy Evolution and Feedback Across Different Environments}
\editors{Thaisa Storchi Bergman et al. eds.}

\begin{document}

\maketitle

\begin{abstract}
The evolution of galaxies at Cosmic Noon (1$<${\it z}$<$3) passed through a dust-obscured phase, during which most stars formed and black holes in galactic nuclei started to shine, which cannot be seen in the optical and UV, but it needs rest frame mid-to-far IR spectroscopy to be unveiled. At these frequencies, dust extinction is minimal and a variety of atomic and molecular transitions, tracing most astrophysical domains, occur. The future IR space telescope mission, SPICA, currently under evaluation for the 5th Medium Size ESA Cosmic Vision Mission, fully redesigned with its 2.5-m mirror cooled down to $T < 8\, \rm{K}$ will perform such observations. SPICA will provide for the first time a 3-dimensional spectroscopic view of the hidden side of star formation and black hole accretion in all environments, from voids to cluster cores over 90\% of cosmic time. Here we outline what SPICA will do in galaxy evolution studies. 




\end{abstract}

\firstsection 
\section{Introduction}
\begin{figure}[t]
\vspace*{-0.2 cm}
\begin{center}
    \includegraphics[width=13cm]{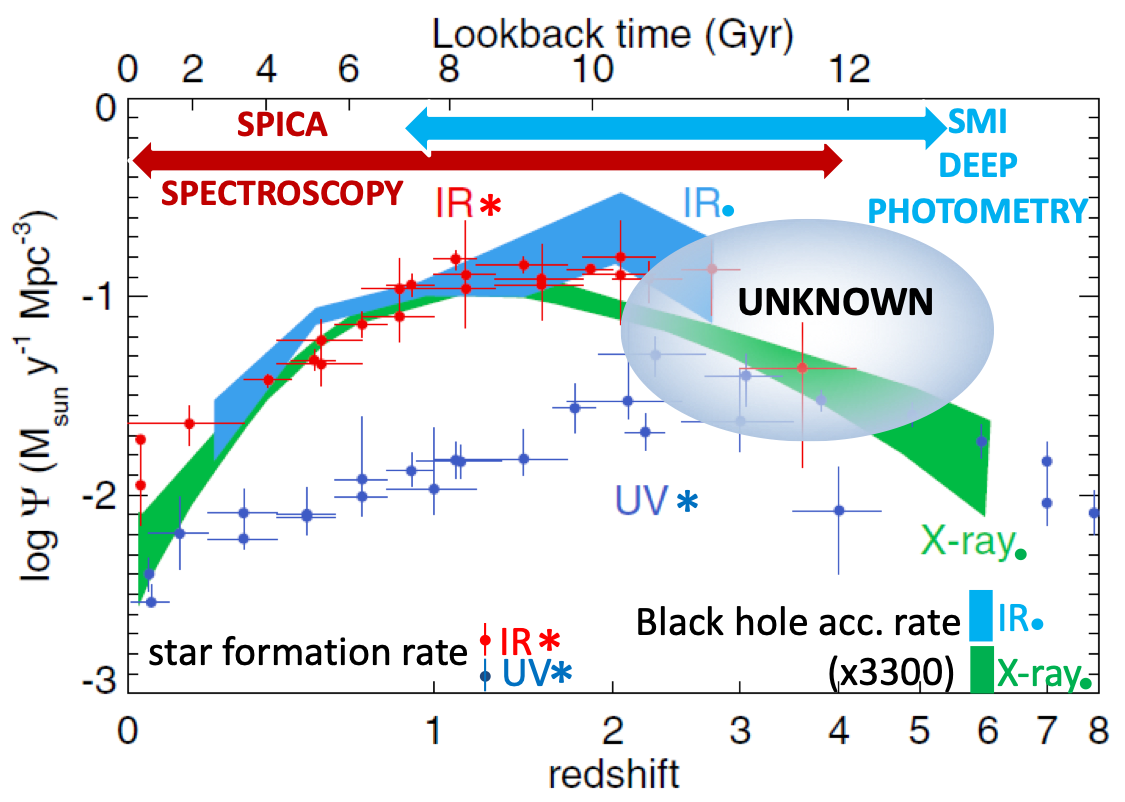}
    \caption{Estimated star-formation rate densities from the far-ultraviolet (blue points) and far-IR (red points) photometric surveys \citep[figure adapted from][]{ma_di14}. The estimated BHAR density, scaled up by a factor of $3300$, is shown for comparison (in green shading from X-rays and light blue from the IR).}\label{fig_madau}
\end{center}
\end{figure}

Most of the activity in galaxy evolution, the formation of stars and supermassive black holes at the center of galaxies, took place more than six billion years ago, with a sharp drop to the present epoch \citep[e.g.,][Fig.\,\ref{fig_madau}]{ma_di14}. Since most of the energy emitted by stars and accreting SMBHs is absorbed and re-emitted by dust, understanding the physics of galaxy evolution requires infrared (IR) observations of large, unbiased galaxy samples spanning a range in luminosity, redshift, environment, and nuclear activity. From {\it Spitzer} and {\it Herschel} photometric surveys the global Star Formation Rate (SFR) and Black Hole Accretion Rate (BHAR) density functions have been {\it estimated} through measurements of the bolometric luminosities of galaxies \citep{leflo05, gru13, del14}. However, such integrated measurements could not separate the contribution due to star formation from that due to BH accretion
\citep[see, e.g.,][]{mul11}. This crucial separation has been attempted so far through modelling of the spectral energy distributions and relied on model-dependent assumptions and local templates, with large uncertainty and degeneracy. On the other hand, determinations from UV \citep[e.g.][]{bou07} and optical spectroscopy \citep[e.g., from the {\it Sloan} Digital Sky Survey, ][]{eis11} track only marginally ($\sim$\,10\%) the total integrated light (Fig.\,\ref{fig_madau}). X-ray analyses of the BHAR, in turn, are based on large extrapolations and possibly miss a large fraction of obscured objects.
Furthermore, the SFR density at $z$ $>$ 2-3 is very uncertain, since it is derived from UV surveys, highly affected by dust extinction. As opposite, through IR emission lines, the contributions from stars and BH accretion can be separated. {\it SPICA} spectroscopy will allow us to directly measure redshifts, SFRs, BHARs, metallicities and dynamical properties of gas and dust in galaxies at lookback times down to about 12 Gyrs. {\it SPICA} spectroscopic observations will allow us for the first time to redraw the SFR rate and BHAR functions (Fig.\,\ref{fig_madau}) in terms of measurements directly linked to the physical properties of the galaxies. 

\begin{figure}[t]
    \includegraphics[width=13cm]{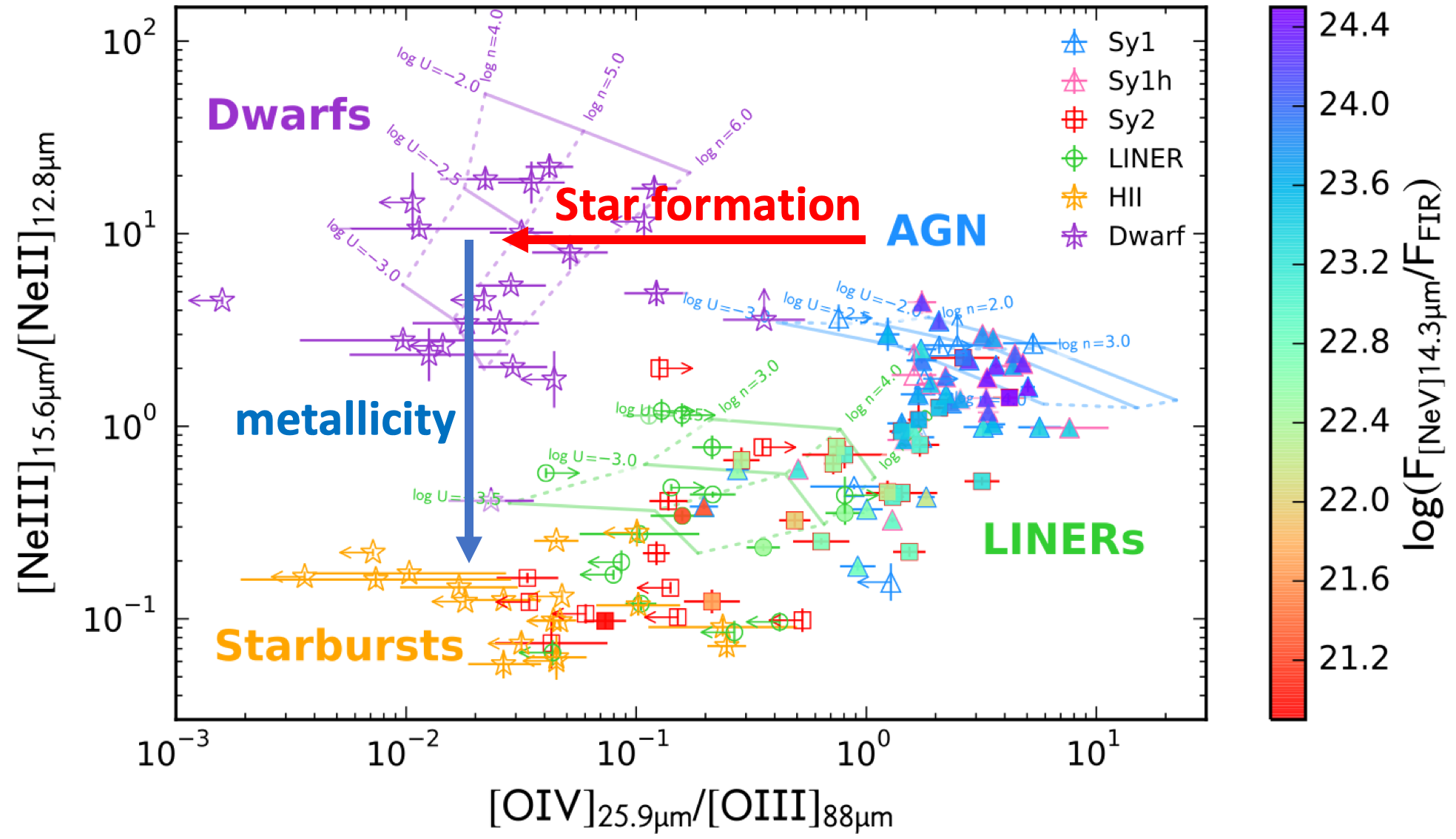}
    \caption{ Observed line ratios of [NeIII]15.6$\mu$m/[NeII]12.8$\mu$m vs. [OIV]26$\mu$m/[OIII]88$\mu$m for AGN, LINER, starburst and dwarf galaxies in the local Universe. These ratios are powerful diagnostic tools: the star formation in galaxies increases from right to left, while the gas metallicity increases from the top to the bottom. For this reason we named this diagram the {\it new BPT diagram} in the IR \citep{fer16}. The active galaxies symbols have been color-coded from blue to red, according to the value of the ratio of the [NeV14.3]$\mu$m line to the FIR continuum as measured with {\it Herschel-PACS} spectra at $\sim$\,160$\mu$m, corresponding the AGN dominance, taken from \citet{fer16}. The solid and dashed lines show standard photoionization models for AGN, LINERS galaxies and dwarf {\it low-metallicity} galaxies. }\label{fig_IR_BPT}
   \vspace*{-0.3cm}
\end{figure}

The mid- to far-IR spectral range hosts a suite of atomic and ionic transitions, covering a wide range of excitation, density and metallicity, directly tracing the physical conditions in galaxies. Ionic fine structure lines (e.g. [NeII], [SIII], [OIII]) probe HII regions around hot young stars, providing a measure of the SFR and the gas density. Lines from highly ionized species (e.g. [OIV], [NeV]) trace the presence of AGN and can measure the BHAR \citep{sm92}. 
Through line ratio diagrams, like the {\it new IR BPT diagram} \citep[][Fig.\,\ref{fig_IR_BPT}]{fer16}, IR spectroscopy can separate the galaxies in terms of both the source of ionization --\,either young stars or AGN excitation\,-- and the gas metallicity, during the dust-obscured era of galaxy evolution ($0.5 < z < 4$). 

\vspace*{-0.6 cm}

\section{SPICA observations of galaxy evolution}

For a complete description of the SPace Infrared telescope for Cosmology and Astrophysics ({\it SPICA}), we refer to \citet{roe18}. We concentrate here on the work that {\it only} an IR observatory such as SPICA will be able to do to unveil galaxy evolution.

\begin{enumerate}
\item	SPICA will provide for the first time a 3-dimensional spectroscopic view of the hidden side of star formation and black hole accretion in all environments, from voids to cluster cores over 90\% of cosmic time. This will lead to a complete  census -- as a function of cosmic time -- of: 

\subitem $\bullet$	the star formation in galaxies from low-mass dust-poor star-forming galaxies to high-mass heavily dust-obscured starbursts, setting a fundamental new benchmark for cosmological models, allowing unambiguous tests of the physics driving star, galaxy and large-scale structure formation. 
\subitem $\bullet$	the black hole accretion in the nuclei of galaxies, including also heavily obscured Compton thick nuclei. 

From these observations, we will be able to derive both the average cosmic star-formation and accretion histories of the Universe as well as the physical conditions of each individual galaxy through their spectra in the large cosmic time interval since about 12 billion years \citep{spi17}. SPICA will allow us to understand the physical origin of the dramatic change in the efficiency of star formation and accretion around cosmic noon, one of the major challenges for theoretical models in present-day cosmology. 

\item SPICA will measure if and how energetic (radiative and mechanical) feedback from starburst and AGN influences galaxy evolution. SPICA will determine the role of active galactic nuclei in quenching star formation in galaxies and address the origin of the low stellar to dark matter halo mass ratio in massive galaxies \citep{gon17}. 

\item SPICA will determine how galaxies build up their metals and dust during the last ten Gyrs of cosmic time and how are these metals recycled in the ISM and injected into the CGM \citep{fer17}. IR line based tracers will allow SPICA to peer through the heavily opaque medium of main-sequence galaxies at the cosmic noon and probe their true chemical ages. Galaxies at 
z $\sim$ 2-3 show extremely low metallicities in optically-based traces, down to 0.8 dex below solar \citep{ono16}, which is in conflict with the large amounts of dust inferred from IR photometry (Fig. 1), suggesting much higher metallicities. SPICA will also probe the first production of dust and metals, assess the presence of hot dust in the most distant galaxies (z$>$6) and determine the physical properties of infrared bright galaxies and quasar hosts at the epoch of re-ionization.
\end{enumerate}

\subsection{SPICA observational strategy to unveil galaxy evolution}\label{strategy}

We briefly show here some of the results of the scientific work that has been done in preparation of the galaxy evolution studies that will performed with the SPICA mission.

In the context of the development of a coherent observing strategy, we have considered a step by step observational sequence, aimed at optimising the measure of basic astrophysical quantities in galaxies through their evolution in the last 12 Gyrs. This sequence can be summarised as follows:

{\it (i)} Deep spectrophotometric and photometric surveys in the mid-IR, using the SPICA Mid-IR Instrument (SMI) of large enough fields (of order 1-10 deg$^2$) down to very faint flux limits (i.e. 3-12$\mu$Jy in the continuum at 34$\mu$m and 2-5$\times$10$^{-20}$ W/m$^2$ in the spectral lines). These observations will be complemented with the B-BOP camera at 70$\mu$m (at a 30$\mu$Jy depth). {\it This survey will provide a 3-dimensional view of galaxy evolution.}

{\it (ii)} Identification of a sample of galaxies based on the outcome of the above surveys, in terms of redshift, mid-IR luminosity, classification (i.e. Starburst-dominated or AGN-dominated), characterisation of the Stellar Masses, Dust Masses and (where possible) Bolometric Luminosities using available multifrequency data.

{\it (iii)} Deep follow-up observations with the grating spectrometers (both SAFARI and SMI at spectral medium resolution) of above defined sample. 
{\it This will provide a complete spectral atlas of galaxies as a function of redshift, luminosity and stellar mass.}

From the spectral atlas of galaxies we will determine their physical properties as a function of cosmic time:  the Star Formation Rate (SFR), the Black Hole Accretion Rate (BHAR), the metal abundances, the outflow/infall occurrence, etc. 

\subsection{Predictions of the SMI ultra-deep survey}

Following the work already presented in \citet{kan17} and \citet{gru17}, we are planning to perform an ultra-deep survey of 1 deg$^2$ with the SMI imaging low-resolution spectrometer \citep{kan18} down to the photometric limit of  3$\mu$Jy in the continuum at 34$\mu$m in  a typical observing time of about $\sim$ 600 hours.

\begin{figure}[h]
\begin{center}
    \includegraphics[width=7cm]{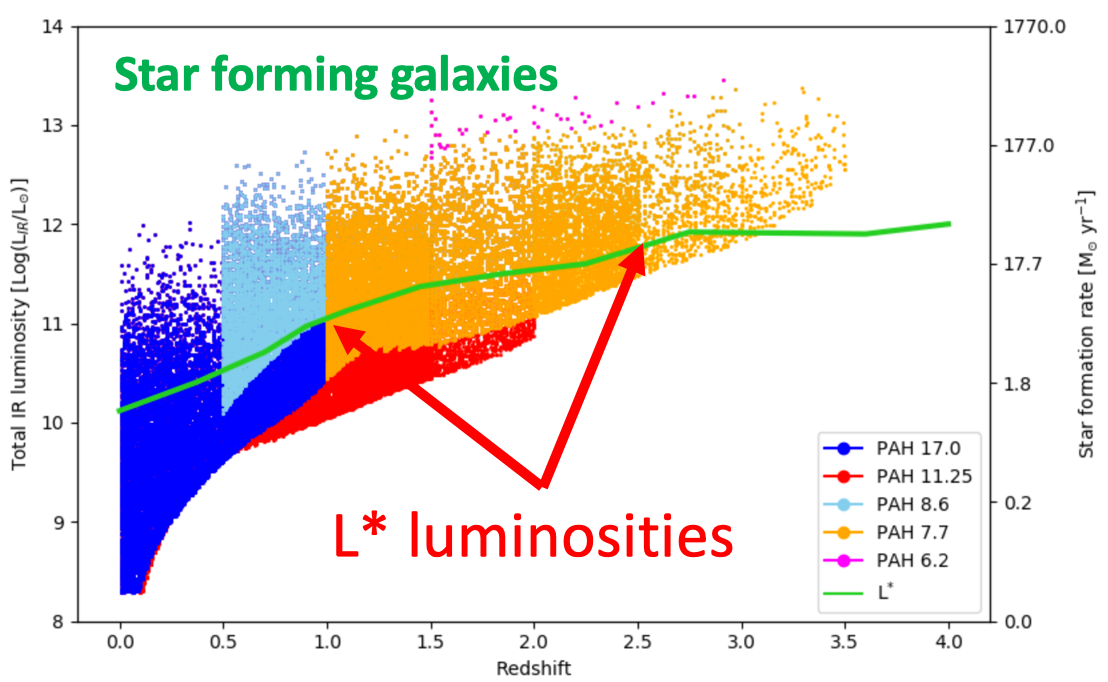}~
    \includegraphics[width=6.7cm]{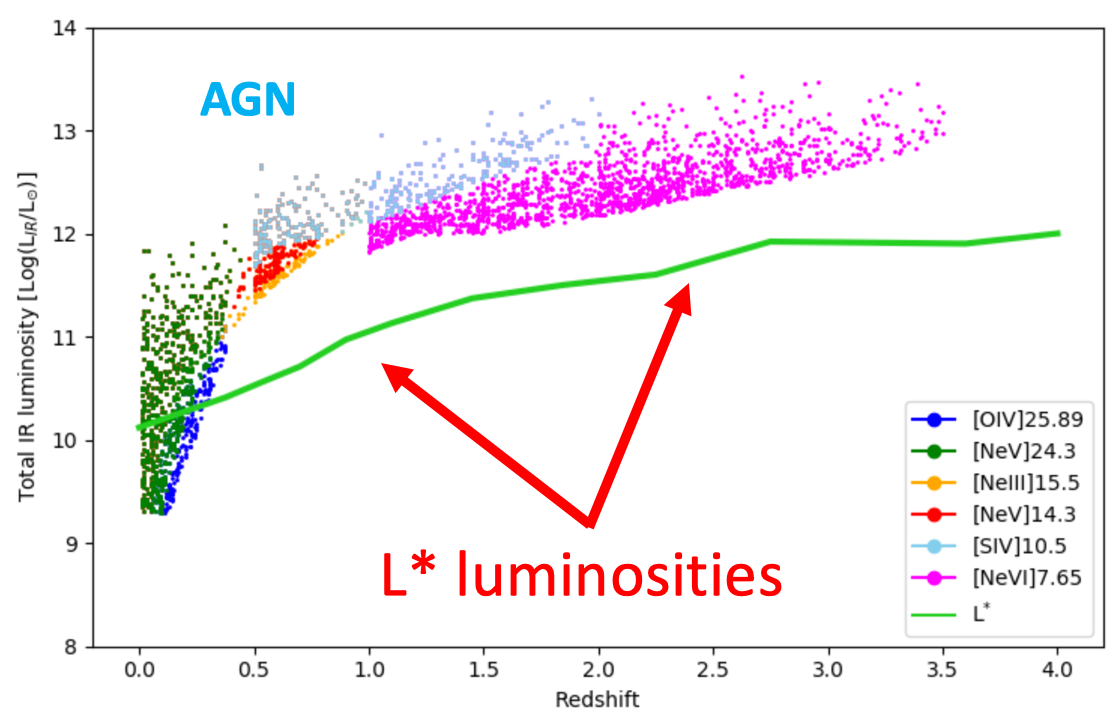}
    \caption{{\bf Left:} Redshift-Luminosity diagram with the simulation of the SMI ultra-deep survey of 1 deg$^2$ showing the Star Forming Galaxies detections in the various PAH spectral features. The green solid line shows the knee of the Luminosity Functions, as a function of redshift. {\bf Right:} Same, but simulated with the AGN detections in high-ionization fine-structure lines.}\label{fig_SMI}
\end{center}
\end{figure}

We show in Fig. \ref{fig_SMI} the predictions of how the SPICA SMI ultra-deep survey of 1 deg$^2$ will fill the luminosity-redshift plane for Star Forming Galaxies (left diagram) and for AGN (right diagram). The Star Forming galaxies are detected through the Polycyclic Aromatic Hydrocarbon (PAH) features at 6.2, 7.7, 8.6, 11.25 amd 17.0 $\mu$m, while the AGN can be detected through a set of high-ionization fine-structure lines, [NeVI]7.65$\mu$m, [SIV]10.5$\mu$m, [NeV]14.3 and 24.3$\mu$m,  [NeIII]15.5$\mu$m, and [OIV]15.9$\mu$m.

We have used the far-IR luminosity functions as derived from {\it Herschel} photometric observations \citep{gru13} and the IR linea and PAH features calibrations to the total IR luminosities from observations in the local Universe. The details of these simulation can be found in Mordini et al (2020, in preparation). 

\subsection{Predictions of SAFARI follow-up pointed spectroscopic  observations}

\begin{figure}[h]
\begin{center}
    \includegraphics[width=7cm]{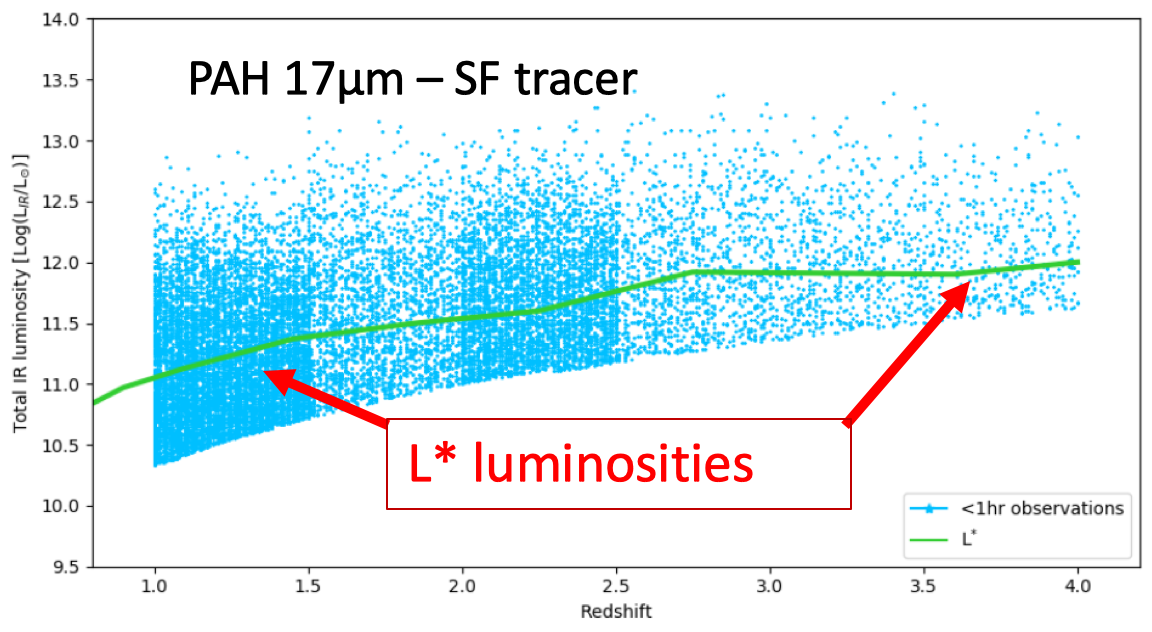}~
    \includegraphics[width=6.7cm]{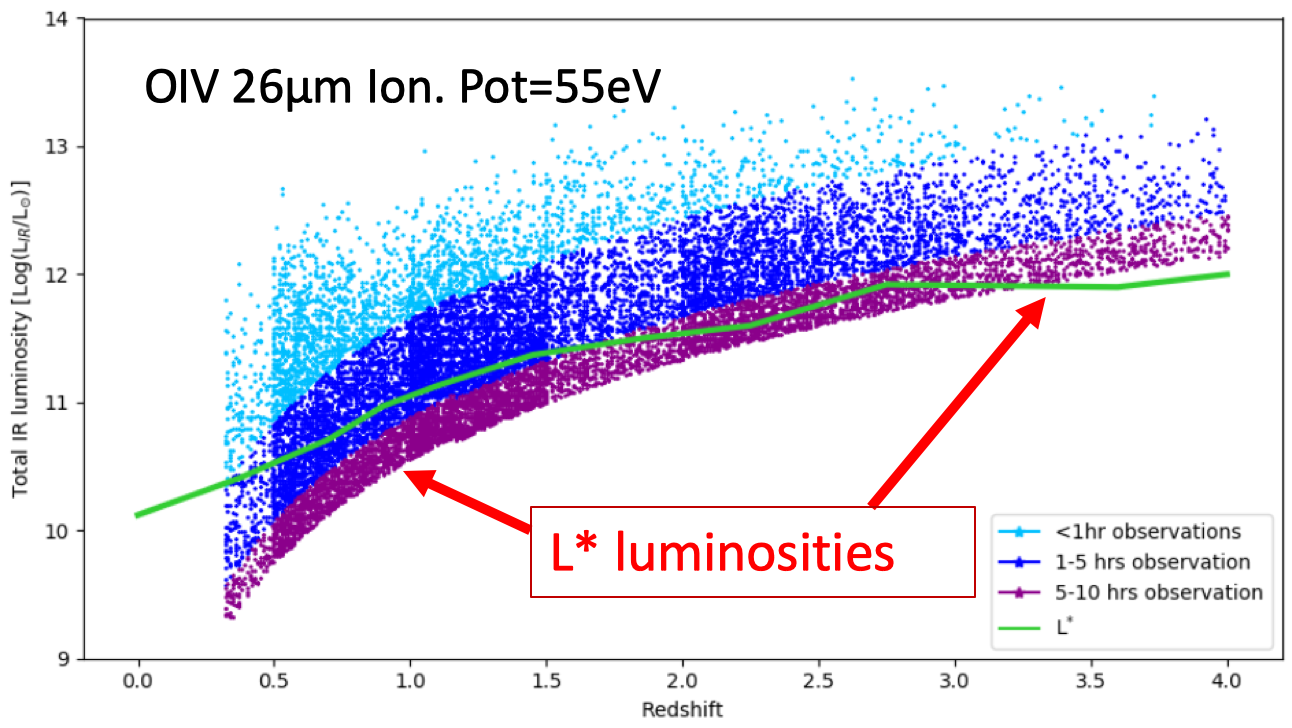}
    \caption{{\bf Left:} Redshift-Luminosity diagram with the simulation of the SAFARI follow-up pointed observations of Star Forming Galaxies with the PAH 17$\mu$m  spectral features. The green solid line shows the knee of the Luminosity Functions, as a function of redshift. {\bf Right:} Same, but simulated with the AGN detections of the [OIV]26$\mu$m fine-structure line.}\label{fig_SAFARI}
\end{center}
\end{figure}

Following step {\it ii)} outlined in Section \ref{strategy}, i.e. the definition of a suitable sample of galaxies, of order 1,000 objects, we have made predictions of the observability of such sample with the SAFARI spectrometer \citep{roe18}. We show these predictions in Fig. \ref{fig_SAFARI}, where, on the left, the luminosity-redshift plane for Star Forming Galaxies through the detection of the Polycyclic Aromatic Hydrocarbon (PAH) feature at 17$\mu$m is shown. Analogously, on the left diagram of Fig. \ref{fig_SAFARI}, the luminosity-redshift plane for AGN is shown, where the detections by SPICA-SAFARI are made in the [OIV]25.9$\mu$m line. We can see from these predictions that SPICA will be able to measure the SFR and the BHAR of galaxies far below the knee of the luminosity functions at each redshift, in particular beyond z$\sim$4 for Star Forming galaxies and up to z$\sim$3.5 for AGN.

\vspace{-0.4cm}

\section*{Acknowledgements}
\textit{
This paper is dedicated to the memory of Bruce Swinyard, who initiated the SPICA project in Europe, but unfortunately died on 22 May 2015 at the age of 52. He was ISO-LWS calibration scientist, Herschel-SPIRE instrument scientist, first European PI of SPICA and first design lead of SAFARI. 
We acknowledge the whole SPICA Collaboration Team, as without its multi-year efforts and work this paper could not have been possible. 
We also thank the SPICA Science Study Team appointed by ESA and the SPICA Galaxy Evolution Working Group.
LS and JAFO acknowledge financial support by the Agenzia Spaziale Italiana (ASI) under the research contract 2018-31-HH.0.}
\vspace*{-0.2cm}

{\small
\begin{multicols}{2}

\end{multicols}
}


\end{document}